\documentclass[preprint,5p,times,twocolumn,compress]{elsarticle}

\usepackage{hyperref}
\usepackage{amsmath, amssymb}
\usepackage{graphicx}
\usepackage{listings}
\usepackage{xcolor}
\usepackage{booktabs}
\usepackage{enumitem}
\usepackage{tabularx}
\usepackage{makecell}
\usepackage{threeparttable}
\usepackage{caption}

\journal{Computer Physics Communications}

\definecolor{codegreen}{rgb}{0,0.6,0}
\definecolor{codegray}{rgb}{0.5,0.5,0.5}
\definecolor{codepurple}{rgb}{0.58,0,0.82}
\definecolor{codefont}{rgb}{0.44,0.34,0.59}
\definecolor{codefont1}{rgb}{0.48, 0.24, 0.62}
\definecolor{codefont2}{rgb}{0.67, 0.22, 0.19}
\definecolor{backcolour}{rgb}{0.965,0.965,0.945}
\definecolor{Brown}{cmyk}{0,0.81,1,0.60}
\definecolor{OliveGreen}{cmyk}{0.64,0,0.95,0.40}
\definecolor{CadetBlue}{cmyk}{0.62,0.57,0.23,0}
\definecolor{lightlightgray}{gray}{0.9}

\usepackage[utf8]{inputenc}
\usepackage{textcomp}        
\DeclareUnicodeCharacter{2605}{\ding{72}}
\usepackage{pifont}


\lstdefinestyle{mystyle}{
    language=Python,
    backgroundcolor=\color{backcolour},   
    commentstyle=\color{codegreen},
    keywordstyle=\color{magenta}\bfseries,
    numberstyle=\tiny\color{codegray},
    stringstyle=\color{codepurple},
    basicstyle=\ttfamily\footnotesize,
    breakatwhitespace=false,         
    breaklines=true,                 
    captionpos=b,                    
    keepspaces=true,                 
    numbersep=5pt,                  
    showspaces=false,                
    showstringspaces=false,
    showtabs=false,                  
    tabsize=2,
}

\renewcommand{\lstinline}[1]{\oldlstinline[basicstyle=\ttfamily]{#1}}

\begin{document}

\begin{frontmatter}

\title{Chromo: A High-Performance Python Interface to Hadronic Event Generators for Collider and Cosmic-Ray Simulations}

\author[ASIOP]{Anatoli Fedynitch\corref{cor1}}
\author[TUDO]{Hans Dembinski}
\author[ASIOP]{Anton Prosekin\corref{cor3}}

\cortext[cor1]{Email: anatoli@gate.sinica.edu.tw}
\cortext[cor3]{Email: antonpr@gate.sinica.edu.tw}

\address[ASIOP]{Institute of Physics, Academia Sinica, Taipei City, 115201, Taiwan}
\address[TUDO]{Department of Physics, TU Dortmund University, D-44227 Dortmund, Germany}

\begin{abstract}
Simulations of hadronic and nuclear interactions are essential in both collider and astroparticle physics. The Chromo package provides a unified Python interface to multiple widely used hadronic event generators, including EPOS, DPMJet, Sibyll, QGSJet, and Pythia. Built on top of their original Fortran and C++ implementations, Chromo offers a zero-overhead abstraction layer suitable for use in Python scripts, Jupyter notebooks, or from the command line, while preserving the performance of direct calls to the generators. It is easy to install via precompiled binary wheels distributed through PyPI, and it integrates well with the Scientific Python ecosystem. Chromo supports event export in HepMC, ROOT, and SVG formats and provides a consistent interface for inspecting, filtering, and modifying particle collision events. This paper describes the architecture, typical use cases, and performance characteristics of Chromo and its role in contemporary astroparticle simulations, such as in the MCEq cascade solver.
\end{abstract}


\end{frontmatter}

\section{Introduction}
Simulations of hadronic, photo-hadronic, and nuclear interactions are central to many problems in high-energy physics. At colliders, event generators are employed to model the underlying event and soft interactions in analyses involving nuclear targets. In cosmic-ray and neutrino physics, these interactions determine air shower development and atmospheric lepton fluxes. Accurate modeling across wide energy and phase space ranges, especially in the forward region, often requires combining multiple event generators and estimating uncertainties through inter-model comparisons.

Despite decades of development, most general-purpose event generators, such as \texttt{Pythia}~\cite{Sjostrand:2006za, Bierlich:2022pfr}, \texttt{DPMJet}~\cite{Roesler:2000he,Engel:1994vs, Fedynitch:2015kcn}, \texttt{QGSJet}~\cite{Kalmykov:1997te, Ostapchenko:2004ss, Ostapchenko:2010vb, Ostapchenko:2024myl}, \texttt{EPOS}~\cite{Pierog:2013ria, Pierog:2023ahq}, \texttt{Sibyll}~\cite{Ahn:2009wx, Riehn:2019jet, Riehn:2024prp}, remain fragmented in interface design, configuration mechanisms, and output formats. This heterogeneity hinders interoperability and efficient model comparison. Existing wrappers like \texttt{CRMC}~\cite{ulrich_ralf_2021_5270381} simplify some aspects but are limited in scope, lack a modern interface, and require substantial setup effort.

To address these challenges, we present the open-source package\footnote{\url{https://github.com/impy-project/chromo}} \texttt{Chromo}\cite{Fedynitch:2025chr}, a Cosmic ray and HadRonic interactiOn MOnte carlo frontend -- implemented in Python. It aims to reduce the friction of using these tools by providing a unified and user-friendly interface for simulation and analysis. \texttt{Chromo} is designed with three primary goals: (1) eliminate the need for platform-specific compilation through binary wheels; (2) enable interactive and script-based usage via a Pythonic API and Jupyter support; and (3) preserve high performance while offering flexible event manipulation, filtering, and export.

\section{Overview}
\texttt{Chromo} acts as a lightweight frontend to a curated collection of hadronic interaction models written in Fortran 77, Fortran 90, or C++. These include legacy models like \texttt{Sibyll-2.1}, widely-used collider-focused tools like \texttt{Pythia 8}, and multipurpose models like \texttt{EPOS-LHC} and \texttt{DPMJet}. Each model is bundled as a Python extension module compiled with \texttt{f2py} or \texttt{pybind11}, with standardized initialization, kinematics configuration, event generation, and output logic.

\subsection{Scientific applications}
The unified interface provided by \texttt{Chromo} allows users to compare interaction models under identical conditions, which is critical for systematic studies. A prominent use case is the generation of particle production matrices used by the MCEq cascade solver to compute atmospheric lepton fluxes. In collider physics, \texttt{Chromo} serves as a drop-in replacement for \texttt{CRMC}, with added capabilities for visualization and custom analysis in Python.

\subsection{Installation and distribution}
For regular users, installation of \texttt{Chromo} is straightforward and requires no compilation from source. The package is distributed through the Python Package Index (PyPI) as platform-specific binary wheels, allowing users to install it simply by running:

\begin{lstlisting}[language=Bash]
pip install chromo
\end{lstlisting}
This eliminates the need to manually compile Fortran or C++ code, which can be a significant barrier for non-expert users. Continuous integration workflows using GitHub Actions and \texttt{cibuildwheel} ensure that validated builds are available for all supported platforms (Linux, macOS, and Windows).

\subsection{Interactive and scripted use}
\texttt{Chromo} is equally suited for command-line use, scripting, and interactive sessions in Jupyter. The API follows Python conventions and provides introspectable classes for kinematics, event generators, and events. Generated events can be streamed, filtered, visualized as SVG graphs, or written to \texttt{HepMC3} and \texttt{ROOT} formats for downstream analysis.

\subsection{Zero-overhead integration}
The internal design leverages memory views and Fortran common blocks to expose event information directly as NumPy arrays, avoiding unnecessary copying. As shown in later performance benchmarks, this allows \texttt{Chromo} to match or even outperform traditional wrappers, especially for fast generators like \texttt{Sibyll}.

\subsection{Basic example}
A typical workflow involves setting up the initial state of a collision using the \lstinline{CenterOfMass} or \lstinline{FixedTarget} classes from the \lstinline{kinematics} module, initializing a model, and streaming events:

\begin{lstlisting}[language=Python]
from chromo.kinematics import CenterOfMass
from chromo.models import EposLHC
from chromo.constants import TeV

collision_kin = CenterOfMass(1 * TeV, "p", "O")
event_generator = EposLHC(collision_kin)

for event in event_generator(100):
    print(event.final_state().pt)
\end{lstlisting}

Users can filter events, export to disk, or visualize them interactively. The event objects expose particle data as NumPy arrays and include metadata for reproducibility.

\section{Example usage}

This section illustrates how Chromo can be used to simulate events, analyze particle properties, and export data, all within a modern Pythonic workflow. The design emphasizes ease of use for both quick interactive exploration and large-scale data production.

\subsection{Basic workflow}

A typical workflow begins by defining the collision kinematics and selecting an event generator. Chromo provides specialized classes to encode frame-specific configurations:

\begin{lstlisting}[language=Python]
from chromo.kinematics import CenterOfMass
from chromo.models import DpmjetIII193
from chromo.constants import TeV

# Define 14 TeV proton-proton collision
kin = CenterOfMass(14 * TeV, "p", "p")
# Initialize an event generator with kinematics
gen = DpmjetIII193(kin)

# Generate 100 events
for event in gen(100):
    # Get particles in final state
    final_state_particles = event.final_state()
    # Process event (e.g. print pT) 
    print(final_state_particles.pt)
\end{lstlisting}

\subsection{Particle filtering and derived quantities}

Each \texttt{event} object provides NumPy views to HEPEVT-style data. Common operations include filtering, histogramming, and calculating derived observables:

\begin{lstlisting}[language=Python]
# Filter final charged particles
charged = (event.status == 1) & (event.charge != 0)
# Transverse momentum
pt = event.pt[charged]
# Pseudorapidity
eta = event.eta[charged]
# Feynman-x
xf = event.xf[charged]
\end{lstlisting}

No data is copied unless explicitly requested. The default interface is optimized for memory locality and allows event filtering via boolean masks.

\subsection{Working with composite targets}

In many applications, such as air shower simulations or fixed-target experiments, the target may consist of a mixture of nuclei. Chromo supports this use case through the \texttt{CompositeTarget} class, which allows users to define a probabilistic mixture of nuclei:

\begin{lstlisting}[language=Python]
from chromo.util import CompositeTarget
from chromo.kinematics import CenterOfMass
from chromo.models import EposLHC
from chromo.constants import TeV

# Define atmospheric air as 78% N, 21% O, 1% Ar
air_components = ("N", 0.78), ("O", 0.21), ("Ar", 0.01)
air = CompositeTarget(air_components)
kin = CenterOfMass(1 * TeV, "p", air)
gen = EposLHC(kin)
\end{lstlisting}

Internally, Chromo samples target nuclei from a multinomial distribution according to their specified weights. To minimize initialization overhead, which can be significant for some generators, Chromo precomputes the number of events to simulate for each target nucleus and processes them in contiguous blocks. This avoids multiple re-initialization of the generator with different nuclear targets.

\subsection{A mini-analysis}

\begin{figure*}[t]
  \centering
  \includegraphics[width=\columnwidth]{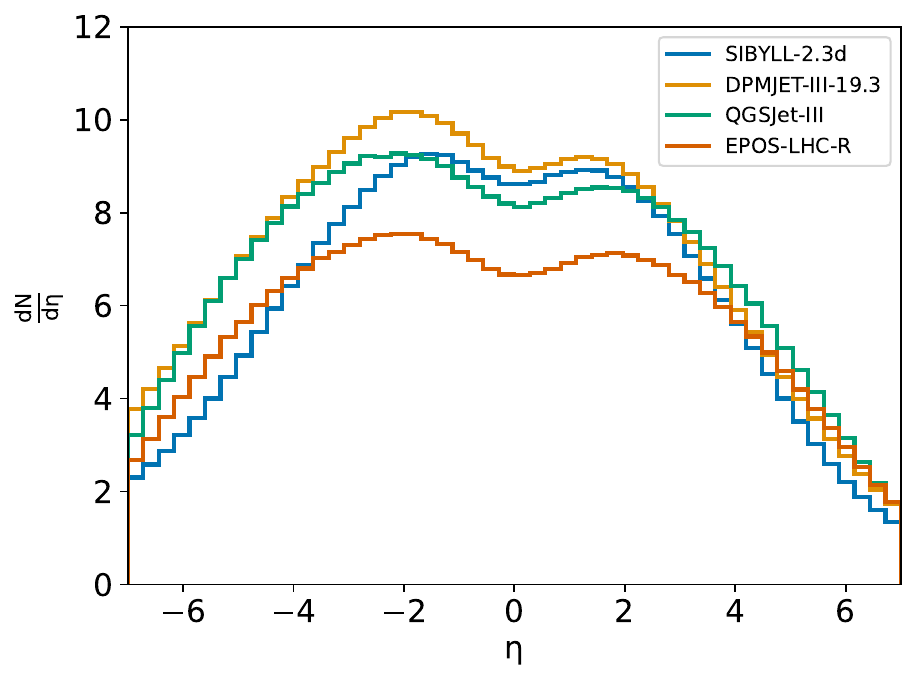}
  \includegraphics[width=\columnwidth]{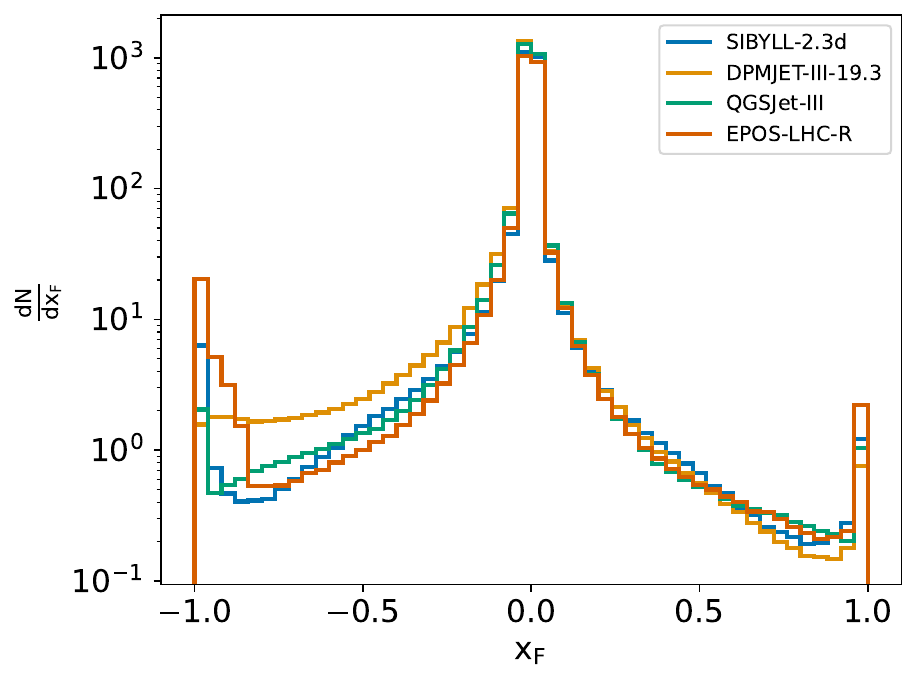}
  \includegraphics[width=\columnwidth]{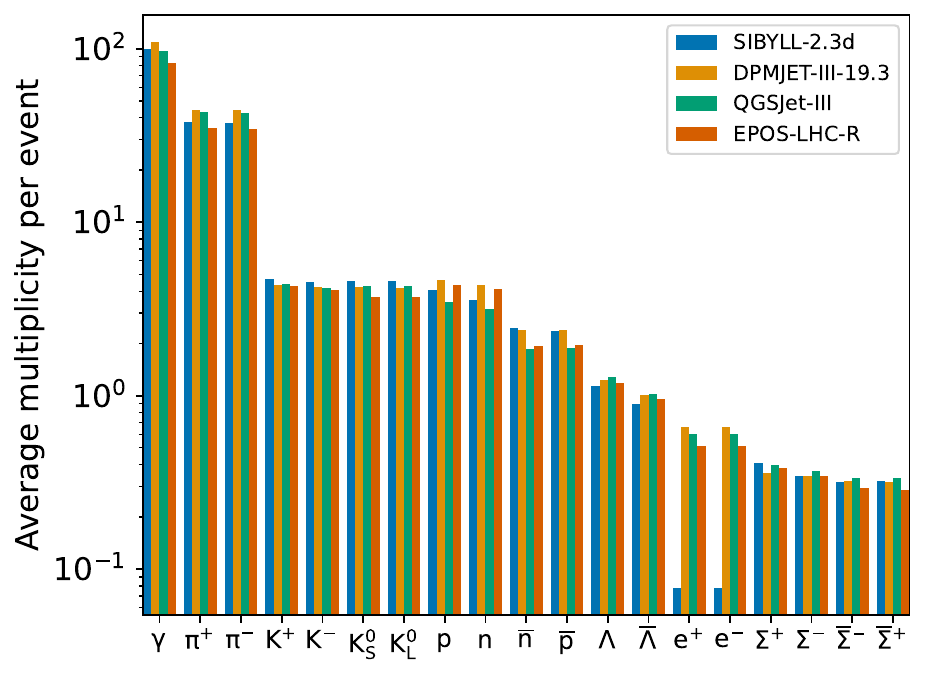}
  \includegraphics[width=\columnwidth]{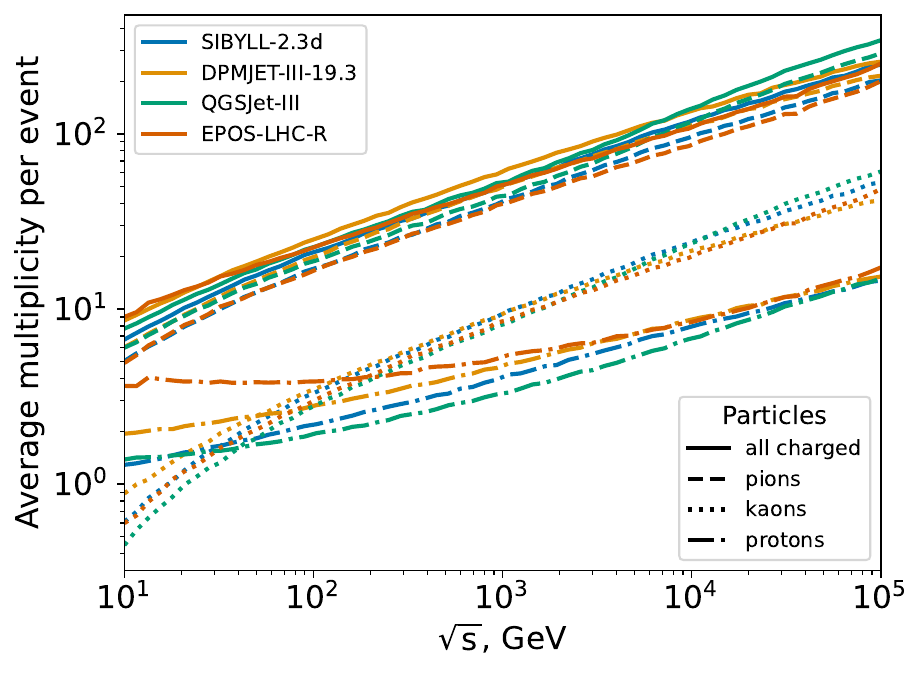}
    \caption{Comparison of Sibyll‑2.3d, DPMJET‑III‑19.3, QGSJet‑III and EPOS‑LHC-R hadronic interaction models for $p+^{16}\mathrm{O}$ collisions at $\sqrt{s}=5\,$TeV. Top left: pseudorapidity distribution $\mathrm{d}N/\mathrm{d}\eta$ of all final charged particles. Top right: Feynman–$x$ distribution $\mathrm{d}N/\mathrm{d}x_F$ of all final charged particles. Bottom left: average multiplicity per event for particle species with multiplicity above a predefined threshold. Bottom right: energy dependence of the average multiplicity per event for all charged particles (solid), pions (dashed), kaons (dotted) and protons (dash‑dotted).}
  \label{fig:histograms_example}
\end{figure*}

The following example demonstrates a mini-analysis: generating events, filtering final-state particles, histogramming key observables, and visualizing the result. It illustrates Chromo's native compatibility with the scientific Python stack.

\begin{lstlisting}[language=Python]
# Highly efficient histogramming library
import boost_histogram as bh
# scikit-hep module for particle properties
from particle import Particle
# Progress bar
from tqdm.auto import tqdm

from chromo.kinematics import CenterOfMass
from chromo.models import Sibyll23d
from chromo.constants import TeV, MeV

# Setup: p + O16 collisions at sqrts = 5 TeV
kin = CenterOfMass(5 * TeV, "p", (16, 8))
gen = Sibyll23d(kin)

# Initialize histograms for Feynman-x and
# pseudorapidity of protons, 
# neutral and charged pions
pid_axis = bh.axis.IntCategory([2212, 111, 211, -211])
hist_xf = bh.Histogram(pid_axis, bh.axis.Regular(50, -1, 1))
hist_eta = bh.Histogram(pid_axis, bh.axis.Regular(50, -7, 7))

# Apply a pT-cut and fill histograms
ptcut = 250 * MeV
nevents = 10000
for event in tqdm(gen(nevents), total=nevents):
    f = event.final_state()
    select = f.pt > ptcut
    hist_xf.fill(f.pid[select], f.xf[select])
    hist_eta.fill(f.pid[select], f.eta[select])

# Plot the pseudorapidity-distribution 
# for charged pions
from matplotlib import pyplot as plt
fig, ax = plt.subplots()
for pid in (211, -211):
    hist_pid = hist_eta[bh.loc(pid), :]
    edges = hist_pid.axes[0].edges
    counts = hist_pid.view()
    particle_name = Particle.from_pdgid(pid).latex_name
    ax.stairs(counts, edges, label=rf"${particle_name}$")

ax.set_xlabel(r"$\eta$")
ax.margins(x=0)
ax.set_ylabel("Counts")
ax.legend()
plt.show()
\end{lstlisting}
This recipe produces histograms similar to those shown in Figure~\ref{fig:histograms_example}.

\subsection{Model switching and parameter scans}
All event generator classes in Chromo inherit from a common base class \texttt{MCRun}, which allows seamless switching between different models. For example, one can loop through models to compare results across them:
\begin{lstlisting}[language=Python]
from chromo.models import Sibyll23d, EposLHC
...
# Initialize generators for each model
# and collect them in dictionary
gens = {model.label : model(kin) for model in [Sibyll23d, EposLHC]}

# Loop through each generator and collect events
for name, gen in gens.items():
    for event in gen(100):
        ...
\end{lstlisting}

Because \texttt{MCRun} defines a writable \texttt{.kinematics} attribute, you can reset collision parameters, such as center‑of‑mass energy or projectile/target nuclei, on the fly without reinitializing the generator:

\begin{lstlisting}[language=Python]
...
# Initialize with maximum energy
gen = Sibyll23d(CenterOfMass(10 * TeV, "p", "p"))

# Scan over different sqrt(s) values
for sqrts in [1, 5, 10]:
    gen.kinematics = CenterOfMass(sqrts * TeV, "p", "p")
    for event in gen(100):
        ...
\end{lstlisting}
Note that while some generators tolerate increasing the energy on the fly, others (e.g., \texttt{Phojet} or the \texttt{DPMJet} family) construct internal lookup tables only up to the energy specified at initialization and will abort if higher energies are requested. To ensure consistent behavior when switching between models or varying the energy, each generator should be initialized with the maximum energy and heaviest nuclei intended for the study.

Each generator model can be initialized only once per Python process. A second instantiation causes Chromo to abort, as the underlying Fortran libraries are dynamically loaded into memory and cannot be unloaded or re-initialized within the same session. Reconstructing a generator class from the same module merely reuses the already loaded shared library. If its initialization routine is invoked again, most generators will fail.

To avoid such conflicts and enable safe concurrent execution, each model should be run in a separate Python process. This can be achieved using Python's \texttt{multiprocessing} module with \texttt{maxtasksperchild=1}, ensuring that each worker process initializes exactly one model and then exits. This allows multiple instances of the same or different generators to run in parallel without shared-library interference:
\begin{lstlisting}[language=Python]
from multiprocessing import Pool, cpu_count
import chromo.models as im
from chromo.kinematics import CenterOfMass
from chromo.constants import TeV

def run_model(model_cls, kin):
    gen = model_cls(kin)
    hist = ...  # Initialize histogram(s)
    for event in gen(100):
        ... # Analyse events and fill histograms
    return hist


# Define models to run
models = [
    im.EposLHC,
    im.Sibyll23d,
    im.QGSJetII04,
    im.Pythia6,
    im.Pythia8,
    ...
]

# Common kinematics object
kin = CenterOfMass(10 * TeV, "p", "p")
args = [(m, kin) for m in models]
n_procs = min(len(args), cpu_count())

# Parallel execution with isolated workers
with Pool(processes=n_procs, maxtasksperchild=1) as pool:
    results = pool.starmap(run_model, args)

# Post-process histogram results
for hist in results:
    ...
\end{lstlisting}

\subsection{Definition of stable particles}
In both air-shower and collider simulations, the definition of ``stable" particles depends on the context and specific analysis goals. Chromo allows this behavior to be explicitly configured through its generator interface.

Each generator provides the methods \texttt{set\_stable(pid)} and \texttt{set\_unstable(pid)}, which allow users to mark particles identified by their PDG ID as either stable (to be included in the final state) or unstable (to be decayed, if supported by the generator):

\begin{lstlisting}[language=Python]
from chromo.models import QGSJetII04
from chromo.kinematics import FixedTarget
from chromo.constants import GeV

kin = FixedTarget(100 * GeV, "He4", "Fe56")
gen = QGSJetII04(kin)

gen.set_stable(111)    # pi0 remains stable
gen.set_unstable(3122) # Lambda0 will decay
...
\end{lstlisting}

Not all generators natively support the decay of unstable particles. In particular, models from the \texttt{QGSJet} family may return certain particles undecayed, even if explicitly marked as unstable. For such cases, Chromo provides an optional fallback mechanism via the \texttt{DecayHandler} class, which uses \texttt{Pythia~8} to decay remaining unstable particles and update the event record. This handler is enabled by default only for \texttt{QGSJet} models, but not for others, as most generators implement their own internal decay logic. If particles marked as unstable remain in the final state, Chromo will issue a warning. In such cases, users can manually enable the decay handler as follows:
\begin{lstlisting}[language=Python]
gen._activate_decay_handler(on=True) # enable
gen._activate_decay_handler(on=False) # disable
\end{lstlisting}

The generator has useful property \texttt{final\_state\_particles} that returns a tuple of PDG IDs considered stable by the generator. This property can also be assigned a new list of PDG IDs to redefine the final state according to the needs of the analysis. For example, to declare all particles with lifetimes longer than a given threshold (e.g., $10^{-18}$ sec) as stable, the following utility function can be used:
\begin{lstlisting}[language=Python]
from chromo.util import select_long_lived
gen.final_state_particles = select_long_lived(1e-18)
\end{lstlisting}
This mechanism provides fine-grained and consistent control over decay behavior across different models and facilitates reproducible event selection criteria.

\subsection{Accessing cross sections}

Chromo provides direct access to total and partial cross sections computed by the event generators. This is useful for normalizing event weights, studying energy-dependent behavior, or validating model consistency against external data.

Each generator implements a \texttt{cross\_section()} method that returns a \texttt{CrossSectionData} object. This object contains fields such as total, inelastic, and elastic cross sections, as well as various diffractive components. All values are expressed in millibarns (mb), where available. If a generator does not provide a particular cross section, the corresponding attribute is set to \texttt{NaN}.

\begin{figure}
  \centering
  \includegraphics[width=\linewidth]{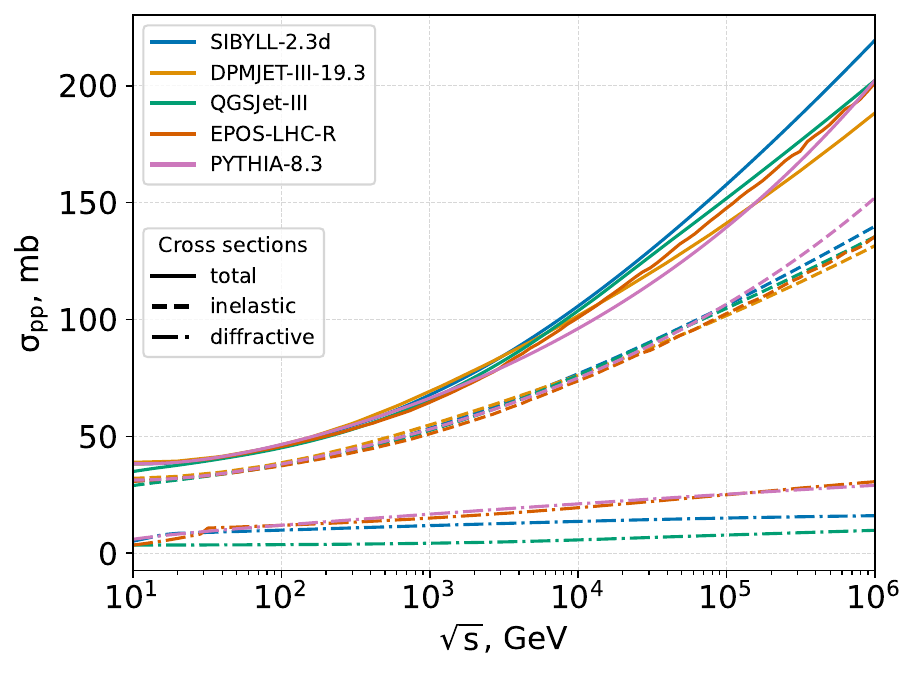}
  \caption{Energy dependence of $pp$ total, elastic, inelastic, and diffractive cross sections for selected models.}
  \label{fig:pp_cross_section}
\end{figure}

\begin{lstlisting}[language=Python]
from chromo.models import EposLHC
from chromo.kinematics import CenterOfMass
from chromo.constants import TeV

gen = EposLHC(CenterOfMass(10 * TeV, "p", "p"))
xs = gen.cross_section()

print("inel =", xs.inelastic, "mb")
print("elas =", xs.elastic, "mb")
print("diff =", xs.diffractive, "mb")
print("total =", xs.total, "mb")
\end{lstlisting}
The cross section API is generator-agnostic. Internally, values are either computed dynamically (e.g., in \texttt{DPMJet}) or taken from pre-tabulated data (e.g., in \texttt{Sibyll} and \texttt{QGSJet}).
Figure~\ref{fig:pp_cross_section} illustrates the energy dependence of various cross section components in \(pp\) collisions for selected models, including total, inelastic, and diffractive contributions.

\subsection{Event serialization}

Generated events can be serialized to HepMC, ROOT (via \texttt{uproot}), or SVG formats using dedicated writer classes. Each writer can be used as a context manager to ensure proper resource handling and automatic file closing. It is possible to combine multiple writers to simultaneously write events in all three available formats:

\begin{lstlisting}[language=Python]
from chromo.writer import Hepmc, Root, Svg
from pathlib import Path as pl

with (Hepmc(pl("output.hepmc3"), gen) as hmc_out,
      Root("output.root", gen) as root_out,
      Svg("output.svg", gen) as svg_out):

    for event in gen(100):
        hmc_out.write(event)
        root_out.write(event)
        svg_out.write(event)
\end{lstlisting}
Events include metadata such as model version, random number generator (RNG) seed, and kinematic configuration to ensure reproducibility.

\subsection{Event inspection and visualization}
\begin{figure*}[t]
  \centering
  \includegraphics[width=\textwidth]{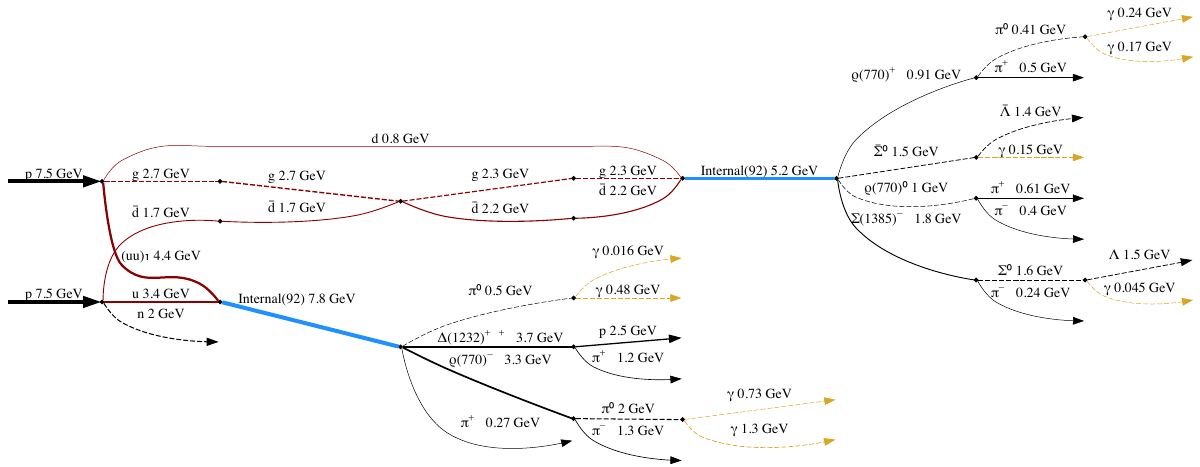}
  \caption{Visualization of a proton-proton collision at $\sqrt{s}=15\,$GeV, generated with Pythia~6 and rendered via \texttt{pyhepmc} using Graphviz.}
  \label{fig:example_event_graph}
\end{figure*}

Printing an event with \texttt{print(event)} shows the raw HepEvt record, a compact Fortran data structure. While efficient, it is not easy to interpret or follow the particle history. Some generators, such as Pythia 6, provide full event histories including intermediate and decayed particles, making them suitable for graphical inspection.

In Jupyter Notebooks, events objects of type \texttt{EventData} are automatically visualized as directed graphs when placed at the end of a cell:
\begin{lstlisting}[language=Python]
# Display a single event in Jupyter
next(iter(gen(1)))
# or, if you already have an event:
event
\end{lstlisting}
An example graph is shown in Figure~\ref{fig:example_event_graph} for an $\sqrt{s} = 15\, $GeV $pp$ collision. Because the output is rendered as rich HTML, one can hover the mouse over particles and vertices to reveal additional tooltip information. Such graphs can be automatically generated for models that expose the complete event generation history including intermediate states and decayed particles.

The automatic visualization is powered by the special method \texttt{\_repr\_html\_()}, and relies on functionality provided by \texttt{pyhepmc}, which uses the optional \texttt{graphviz} package. To further manipulate or customize the visual output, one can use the full \texttt{pyhepmc} API directly. The same visualization backend is also used by the \texttt{Svg} writer, which exports event graphs to SVG files for use outside of Jupyter.

\subsection{Using the command-line interface}

Chromo includes a command-line interface (CLI) for running simulations without writing Python code. This interface is particularly useful for scripted workflows and batch production. A typical command looks like this:

\begin{lstlisting}[language=bash]
chromo -n 1000 -m sibyll23d -S 5000 -i p -I O -f events.hepmc
\end{lstlisting}

\noindent This generates 1000 proton–oxygen collisions at $\sqrt{s} = 5\,\mathrm{TeV}$ using the \texttt{Sibyll\allowbreak -2.3d} model and writes the output in HepMC3 format. 

\vspace{1em}
\noindent\textit{Frequently used CLI options:}\hspace{1em}
\begin{itemize}[itemsep=2pt,parsep=0pt,topsep=2pt]
    \item \texttt{-n}, \texttt{--number} -- number of events to generate
    \item \texttt{-m}, \texttt{--model} -- interaction model (tolerant string match)
    \item \texttt{-S}, \texttt{--sqrts} -- center-of-mass energy $\sqrt{s}$ in GeV
    \item \texttt{-i}, \texttt{--projectile-id} -- projectile (e.g., \texttt{p}, \texttt{pi+}, PDG code)
    \item \texttt{-I}, \texttt{--target-id} -- target (e.g., \texttt{O}, \texttt{N}, \texttt{Pb})
    \item \texttt{-f}, \texttt{--out} -- output file name (format is inferred from extension)
    \item \texttt{-o}, \texttt{--output} -- explicit output format (\texttt{hepmc}, \texttt{hepmc:gz}, \texttt{root}, \texttt{root:vertex}, \texttt{svg} (default is \texttt{hepmc}))
    \item \texttt{-s}, \texttt{--seed} -- random seed (0 means random seed)
    \item \texttt{-h}, \texttt{--help} -- show help message and exit
\end{itemize}

Internally, the CLI constructs the appropriate kinematic configuration, initializes the selected model, and writes events using the same writer backend as the Python API. Model-specific parameters can also be customized using a Python-based configuration file via \texttt{--config}. The CLI mimics the behavior of \texttt{CRMC} \cite{ulrich_ralf_2021_5270381} to facilitate compatibility between these two tools, and to enable deployments in production environments.

\section{Software architecture}
\begin{figure*}
  \centering
  \includegraphics[width=0.9\textwidth]{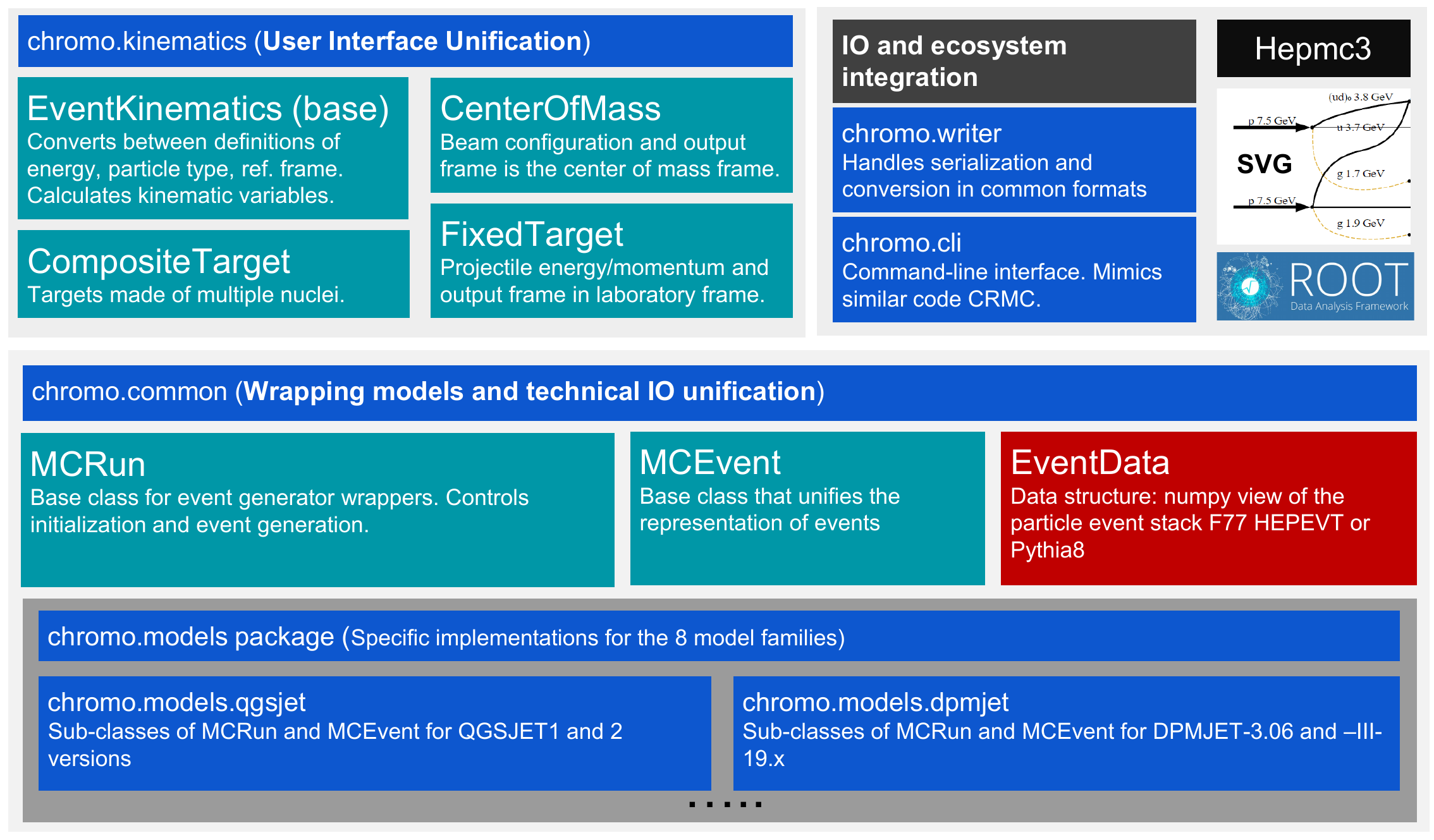}
  \caption{High-level overview of the \texttt{Chromo} architecture through its Python API.}
  \label{fig:program_structure}
\end{figure*}

Chromo is built around three major abstractions: \texttt{Event\allowbreak Kinematics}, \texttt{MCRun}, and \texttt{MCEvent}. The architecture is illustrated in Figure~\ref{fig:program_structure}.

\subsection{Kinematics module}
The \lstinline{kinematics} module provides classes and functions for specifying and manipulating with the initial state of particle interactions in a structured way. The base class \texttt{Event\allowbreak Kinematics\allowbreak Base} stores all relevant information such as incoming particle IDs, energy, momentum, and frame type. 

The two basic specializations are \texttt{Event\allowbreak Kinematics\allowbreak With\allowbreak Restframe} and \texttt{Event\allowbreak Kinematics\allowbreak Massless} where the latter handles the case of collisions between massless particles like photons. The \texttt{beam} argument in these two generic classes expects a pair of arrays of ($p_\mu^\text{particle1}$, $p_\mu^\text{particle2}$), however boosts are restricted to the z-direction except in the case of the \texttt{PHOJET} family of generators. Each kinematics object is frame-aware and supports automatic transformation of event four-vectors, making analysis and output formats consistent using the native generator versions for boosts where possible. The two convenience classes \texttt{CenterOf\allowbreak Mass} and \texttt{Fixed\allowbreak Target} specialize the interface to most popular scenarios and restrict energy and frame parameters to suitable forms.

\subsection{Middle layer, event wrapping, and data handling}

Most Fortran-based event generators store their output in HEPEVT-style common blocks or provide interfaces for conversion into that format. In Chromo, generators from the \texttt{QGSJet} and \texttt{Sibyll} families, for example, use a dedicated Fortran middle layer to convert their internal event records into HEPEVT. This layer also offers a convenient extension point for additional customization in a compiled language.

The Python bindings for each generator are created using \texttt{numpy.f2py}, which exposes selected Fortran subroutines and makes common block memory directly accessible as NumPy arrays. This enables efficient zero-copy access to particle stacks, including the HEPEVT structure. Chromo reads from these shared-memory regions via NumPy views, ensuring that data is only copied when explicitly requested.

The \lstinline{EventData} class wraps these arrays and provides a high-level, NumPy-compatible interface. It supports slicing, filtering, and derived kinematic quantities such as transverse momentum and rapidity, all implemented using vectorized operations for performance and clarity.

When users follow idiomatic NumPy practices, the Python overhead in Chromo is negligible. For example, selecting charged pions with $p_T > 0.5$\,GeV and $|\eta| < 2.5$ can be expressed compactly and efficiently:

\begin{lstlisting}[language=Python]
for event in gen(nevents):
    event = event.final_state_charged()
    selection = (event.pt > 0.5 * GeV) & \
                (np.abs(event.eta) < 2.5) & \
                (np.abs(event.pid) == 211)
    n_pi = np.sum(selection)
\end{lstlisting}

By contrast, using patterns inspired by compiled languages, such as explicit loops over particles, incurs unnecessary overhead and should be avoided:

\begin{lstlisting}[language=Python]
for event in gen(nevents):
    for i in range(len(event)):
        if event.charge[i] == 0:
            continue
        if (event.pt[i] > 0.5 * GeV and 
            abs(event.eta[i]) < 2.5 and
            abs(event.pid[i]) == 211):
            n_pi += 1
    # --> Avoid
\end{lstlisting}

Even in cases where copies of event data are unavoidable (e.g., due to fancy indexing), the overhead is typically small compared to the cost of generating events.

\subsection{Custom \texttt{pybind11} interface for Pythia~8}

In analogy to the Fortran-based middle layer, Chromo includes a custom Python wrapper for the C++-based \texttt{Pythia~8} event generator, implemented using \texttt{pybind11}.\footnote{See \url{https://github.com/pybind/pybind11}} Unlike the official \texttt{Pythia~8} Python bindings, which expose particle information via per-particle accessors, our implementation provides direct NumPy access to entire particle arrays, including four-momenta and auxiliary attributes. This design enables fully vectorized event processing in Python and avoids the overhead associated with Python-level loops and repeated function calls.

Internally, the C++ event data structure follows a layout analogous to HEPEVT, with raw pointer access to the contiguous particle stack. This allows the full event to be transferred to Python in a single call, without copying individual particles or fields.

One of the distinctive features of \texttt{Pythia~8} is its flexible configuration system, which accepts a list of key-value strings. Chromo exposes this interface directly via the \texttt{config} keyword. This covers a wide range of \texttt{Pythia~8} use cases without modifying the C++ interface. By default, Chromo sets \lstinline{SoftQCD:inelastic = on} to match the ``minimum bias'' behavior of the other generators.

To specify a custom configuration, the following pattern can be used:

\begin{lstlisting}[language=Python]
from chromo.models import Pythia8
from chromo.kinematics import CenterOfMass
from chromo.constants import GeV

# Example: e+e- collisions at LEP energies
kin = CenterOfMass(91.2 * GeV, "e+", "e-")

gen = Pythia8(kin, config=[
    "WeakSingleBoson:ffbar2gmZ = on",
    "Print:quiet = off",  # Enable diagnostic output
])
for event in gen(100):
    ...
\end{lstlisting}

This interface allows users to reuse examples and configurations directly from the official \texttt{Pythia~8} documentation. However, the current implementation is not yet a complete replacement for the native interface. Additionally, since \texttt{Pythia~8} does not validate configuration keys or argument types, incorrect settings may lead to segmentation faults that originate in the \texttt{Pythia~8} backend and are not caught by Chromo.

\subsection{Random number generators and seeds}

Random number generators (RNGs) are a core component of all event generators, directly affecting reproducibility and statistical properties of simulations. Most Fortran-based generators bundled with Chromo rely on the RANMAR algorithm~\cite{James:1988vf}, distributed via CERNLIB~\cite{CERNLIB:450356}. In contrast, \texttt{Pythia8} employs its own Mersenne Twister implementation~\cite{mersenne_twister}.

To ensure consistent and reproducible behavior across models, Chromo overrides each generator’s internal RNG with a shared interface to \texttt{numpy.random.Generator}, using the \texttt{PCG-64} backend~\cite{oneill:pcg2014}. This enables transparent seeding and state serialization, including in workflows where random numbers are consumed outside of the Fortran/C++ code, such as when sampling over \texttt{CompositeTarget}s in Python space. The serialization of the RNG state allows for the exact reproduction of events or for continuation from a specific point without accessing the often cryptic interfaces of the original libraries.

\subsection{Writers and exporters}
Output writers are available for \texttt{HepMC3}, \texttt{ROOT} (via \texttt{uproot}), and SVG. Each writer implements a simple interface:
\begin{lstlisting}[language=Python]
from chromo.writer import HepMC

with HepMC("events.hepmc3") as out:
    for event in gen(1000):
        out.write(event)
\end{lstlisting}
Writers can be used inside Python scripts or through the CLI frontend, which mimics \texttt{CRMC}'s behavior while offering a more portable and transparent configuration mechanism.

\subsection{Supported interaction models}
Chromo supports a wide range of hadronic interaction models, covering hadron--nucleon ($hN$), hadron--nucleus ($hA$), nucleus--nucleus ($AA$), photon--nucleon ($\gamma N$), photon--photon ($\gamma\gamma$), and electron--positron ($ee$) collisions. Table~\ref{tab:models} summarizes each model's projectile/target coverage, notable limitations, and event generation performance relative to PYTHIA~8 for 14~TeV proton--proton collisions.

\captionsetup{skip=7pt}
\begin{table*}
  \centering
  \caption{
  Supported interaction models, their projectile/target coverage, and normalized event generation performance. 
  The last column shows the number of events per second relative to PYTHIA~8 for proton--proton collisions at 14~TeV. Note that for \mbox{SOPHIA~2.0} $\gamma$--proton interactions are used, and for \mbox{UrQMD~3.4} collisions at 10~TeV are simulated.
  Citations refer to the corresponding publications or technical descriptions (see Chromo \href{https://github.com/impy-project/chromo?tab=readme-ov-file}{README} for references).}
  \label{tab:models}

  \begin{threeparttable}
  \begin{tabular}{llc}
    \toprule
    \textbf{Interaction Model} & \textbf{Supported proj/targ} &  
    \makecell{\textbf{Normalized} \\ \textbf{performance}} \\
    \midrule
    DPMJET-III 3.0.7 \& PHOJET 1.12-36~\cite{Roesler:2000he,Engel:1994vs} & $hN$, $\gamma\gamma$, $\gamma N$, $hA$, $\gamma A$, $AA$ & 2.1 \& 3.1\\
    DPMJET-III \& PHOJET 19.1/19.3~\cite{Fedynitch:2015kcn} & $hN$, $\gamma\gamma$, $\gamma N$, $hA$, $\gamma A$, $AA$ & 1.9 \& 2.9\\
    EPOS-LHC~\cite{Pierog:2013ria} & $hN$, $hA$, $AA$ & 0.18 \\
    EPOS-LHC-R~\cite{Pierog:2023ahq} & $hN$, $hA$, $AA$ & 0.016 \\
    PYTHIA 6.4~\cite{Sjostrand:2006za} & $hN$, $ee$, $\gamma\gamma$, $\gamma N$ & 2.3\\
    PYTHIA 8.3\tnote{a}~\cite{Bierlich:2022pfr} & $hN$, $ee$, $\gamma\gamma$, $\gamma N$, $hA$, $AA$ & 1.0\\
    QGSJet-01~\cite{Kalmykov:1997te} & $hN$, $hA$, $AA$ & 2.9\\
    QGSJet-II-03~\cite{Ostapchenko:2004ss} & $hN$, $hA$, $AA$ & 0.7\\
    QGSJet-II-04~\cite{Ostapchenko:2010vb} & $hN$, $hA$, $AA$ & 1.1\\
    QGSJet-III~\cite{Ostapchenko:2024myl} & $hN$, $hA$, $AA$ & 0.2\\
    SIBYLL-2.1~\cite{Ahn:2009wx} & $hN$, $hA$ ($A \leq 20$) & 3.8\\
    SIBYLL-2.3\tnote{b}~\cite{Riehn:2019jet} & $hN$, $hA$ ($A \leq 20$) & 3.8\\
    SIBYLL$^\bigstar{}$\tnote{c}~\cite{Riehn:2024prp} & $hN$, $hA$ ($A \leq 20$) & 3.9\\
    SOPHIA 2.0~\cite{Mucke:1999yb} & $\gamma N$ & 6.3\\
    UrQMD 3.4\tnote{a}~\cite{Bass:1998ca,Bleicher:1999xi} & $hN$, $hA$, $AA$ & 0.17\\
    \bottomrule
  \end{tabular}

    \vspace{2pt}
  \begin{center}
    \footnotesize
    $h$ = hadron, $N$ = nucleon (p or n), $A$ = nucleus, $\gamma$ = photon, $e$ = electron/positron
  \end{center}

  \begin{tablenotes}[flushleft]
      \footnotesize
      \item[a] Not available on Windows.
      \item[b] Includes versions 2.3/2.3c/2.3d/2.3e.
      \item[c] Based on 2.3e.
  \end{tablenotes}
  \end{threeparttable}
  
\end{table*}

\section{Performance}

\begin{figure}
  \centering
  \includegraphics[width=\linewidth]{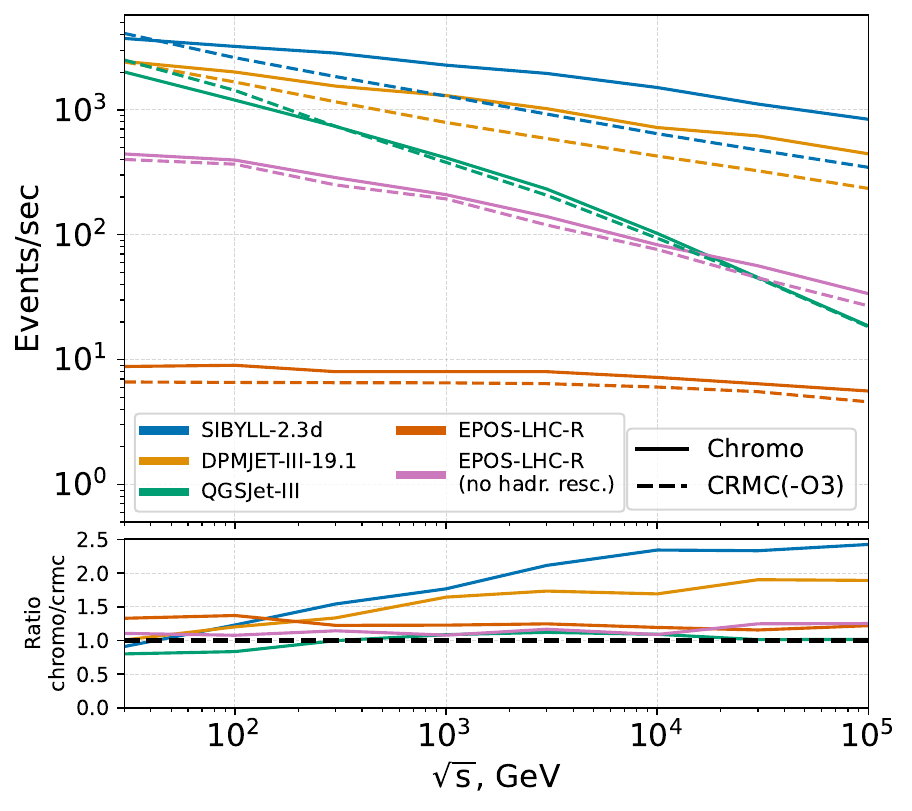}
  \caption{Top: Event generation rates (events/sec) of $pp$ collisions with centre‐of‐mass energy $\sqrt{s}$ for various hadronic interaction models: SIBYLL-2.3d (SIBYLL-2.3e for CRMC), DPMJET-III-19.1, QGSJET-III, EPOS-LHC-R and EPOS-LHC-R without hadronic rescattering. In each case, solid lines denote Chromo and dashed lines CRMC. Bottom: Ratio of Chromo to CRMC event rates for each model.}

  \label{fig:benchmark_result}
\end{figure}

We benchmark Chromo against CRMC by generating proton–proton events over a wide range of center-of-mass energies using several hadronic interaction models: SIBYLL-2.3d (2.3e in CRMC), DPMJET-III-19.1, QGSJet-III, and EPOS-LHCR (with and without hadronic rescattering). Across all energies and models, Chromo matches or exceeds CRMC’s event generation rates (see Fig.~\ref{fig:benchmark_result}). Note that CRMC is compiled in Release mode at \lstinline{-O3} optimization level (instead of the default \lstinline{-O0}) to match Chromo's defaults. These results demonstrate that a high-level interface implemented in Python can deliver competitive performance when carefully designed bindings and memory sharing are employed.

\section{Validation and testing}

Chromo is validated through an extensive test suite comprised of almost 2,000 unit tests. These tests are executed via continuous integration (CI) using GitHub Actions across all supported platforms including Linux, macOS, and Windows and for all Python versions from 3.9 onward. Because event generators incorporate random number generation and floating-point arithmetic, bitwise identical results cannot be expected across architectures or platforms. To address this, Chromo employs probabilistic tests that compare statistical properties of generated events against known reference distributions. The tests verify model correctness while tolerating minor numerical variation. All changes to the codebase are automatically validated in CI to ensure platform-independent consistency, API stability, and reproducibility.

\section{Conclusion and outlook}

We have presented \texttt{Chromo}, a unified Python interface to a comprehensive suite of hadronic interaction event generators, including \texttt{EPOS}, \texttt{DPMJet}, \texttt{Sibyll}, \texttt{QGSJet}, and \texttt{Pythia}. Through careful design of zero-copy bindings, efficient common-block access, and integration with the Scientific Python ecosystem, Chromo achieves performance comparable to or exceeding that of existing wrappers such as CRMC, while offering a high-level, user-friendly API. We have demonstrated a variety of use cases and shown how the package simplifies model comparison, parameter scans, and interactive analysis in Jupyter notebooks.

Chromo is distributed as easy-to-install binary packages, removing the need for manual compilation or dependency management. It addresses longstanding fragmentation in event generator interfaces, configuration styles, and data formats through a unified, high-level API and shared data structures. Its interface enables uniform handling of setup, generation, filtering, and export, reducing boilerplate and lowering the entry barrier for new users. Rigorous probabilistic unit tests and continuous integration ensure reproducibility and cross-platform stability.

Looking ahead, \texttt{Chromo} can continue to evolve in several directions to better support the particle and astroparticle physics communities. Expanding the range of supported event generators and data formats will further increase its utility, especially through the inclusion of specialized models and additional types of particle interactions.

\section*{Acknowledgements}
We thank the developers of the Fortran and C++ event generator codes for their support in interface development. We also gratefully acknowledge the invaluable contributions of early adopters, including Felix Riehn, Keito Watanabe, Sonia El Hedri, Tetiana Kozynets, Dennis Soldin, and members of the LHCb collaboration. AF and AP acknowledge support from Academia Sinica (Grant No.~AS-GCS-113-M04) and the National Science and Technology Council (Grant No.~113-2112-M-001-060-MY3).

\bibliographystyle{elsarticle-num}
\bibliography{references}

\end{document}